# Vector Graph-Based Repository Understanding for Issue-Driven File Retrieval

Authors: Kostiantyn Bevziuk, Andrii Fatula, Svetozar Lashin

Yaroslav Opanasenko, Anna Tukhtarova, Ashok Jallepalli

Pradeepkumar Sharma, Hritvik Shrivastava

Affiliation: Persistent

**Abstract**

We present a repository decomposition system that converts large software repositories into a vectorized knowledge graph which mirrors project architectural and semantic structure, capturing semantic relationships and allowing a significant level of automatization of further repository development. The graph encodes syntactic relations such as containment, implementation, references, calls, and inheritance, and augments nodes with LLM-derived summaries and vector embeddings. A hybrid retrieval pipeline combines semantic retrieval with graph-aware expansion, and an LLM-based assistant formulates constrained, read-only graph requests and produces human-oriented explanations.

**Keywords:** Dynamic knowledge graph; Source code search; RAG-graph; LLM

# 1 Introduction

A big part of the high-level programming repositories used by software developers over the world every day have more than 2000 files; the largest files from those repositories could have 5,000 or more lines of code — those scales exceed LLM context window size by magnitude of 1,000–10,000 times. Based on such limitations, application of highly productive and smart LLM to large code bases becomes a challenge which stays at the front door of automatization of the software development process [8, 3].

Keeping this in mind, we developed a solution that allows us to apply a certain level of automatization and simplification of the development of large code bases for software developers. The most complete and main task that our system is capable of is automatic bug-fix and feature addition / enhancement using only a short user description. The given task can be split into two smaller tasks:

(1) retrieval of relevant source code repository files to a Natural Language (NL) user query / issue description;

(2) applying changes to a set of files selected in Step #1.

In our paper, we present a solution to the first part of the task and explain our own approach of solving given tasks using LLM and the concept of a Dynamic Knowledge Graph, in which we will dive deeper in the next section. Besides the fact that our system is also meant to solve the second part of the main automation problem of large code bases, we suggest reasonable ways to effectively solve the first part of the problem.

Large codebases are intrinsically structured:

(1) directories contain files, files implement classes and functions,

(2) functions call other functions, and classes inherit behavior, etc.

Treating these artifacts and relations as first-class entities enables searches and analyses that are sensitive to both semantics (what code does) and structure (how code is connected) [13, 10]. Figure 1 illustrates how many complex connections between repository components are present in most large code bases, sampling only a small part of the repository structure and connections. The graph contains about 51000 nodes and represents the Apache Airflow repository. In large codebases, the number of nodes could have 6 figures and the number of connections between those nodes could be 10 or more times larger.

We analyze a code repository, breaking it into a structural and semantic map: a repository knowledge graph. Using this graph allows automating retrieval and provides dynamic, context-aware visualizations. In practice, our method leverages modern large language models to help developers work with large codebases, reducing manual effort and increasing automation [1, 2, 14].

## 1.1 Terminology and Definitions

Knowledge Graph (KG) — a graph that contains structured and organized information on the objects, as well as embeddings and connections representing a repository.

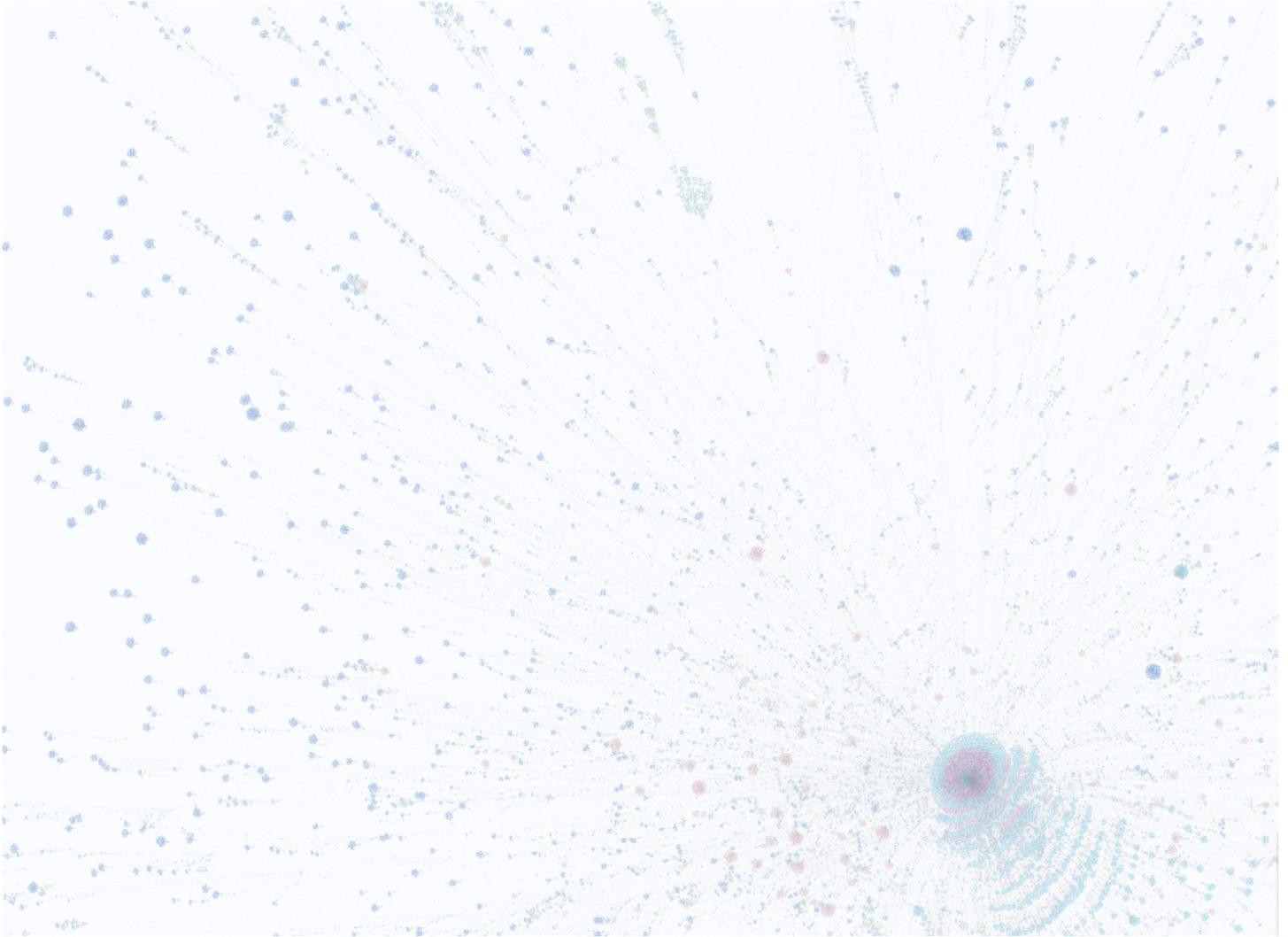

Figure 1: Apache Airflow Dynamic Knowledge Graph screenshot

Dynamic Knowledge Graph (DKG) — a KG that changes over time with every new commit to a repository.

Semantic search (SS) — search among text documents using their vector embeddings generated by LLM.

Search Relevant (SR) — our algorithm that allows retrieval of all objects related to a user query. It is based on semantic search enriched with graph traversal using relations.

## 2 Methodology

In this section, we first present results evaluation metrics, followed by the overall structure of KG and its search scheme, which consists of several steps and approaches. Finally, we describe its inference process and result metrics.

### 2.1 Evaluation Metrics

It is essential to design a fair evaluation protocol to measure the effectiveness of relevant file retrieval. We need an automated framework that can generate many experiments and assess — across multiple repositories — whether the files returned are relevant to a given user query.

We evaluate retrieval quality using GitHub issues that were subsequently addressed by pull requests (PR); each issue text is used as input and the set of files modified in the corresponding PR is taken as ground truth. The evaluation suite spans heterogeneous repositories (services, libraries, crypto middleware, CLI tools, and monorepos) to exercise both semantic and structural aspects of retrieval [16].

As you can imagine, it was very hard to obtain a sufficient number of test cases manually, so we decided to create an automatic test cases generation server. It allowed us to specify repository, branch and datetime for test cases generation.

For evaluation of our approaches, we decided to choose the following metrics: recall, precision, F-beta-score — Fβ@k, recall at K — recall@k, precision at K — precision@k, Percentage found. Our primary metric is median recall per repository, reflecting the goal of maximizing the coverage of relevant files. Precision and other metrics are treated as secondary objectives and will be improved in subsequent iterations.

Recall, also known as sensitivity, is the fraction of relevant instances that were retrieved. It answers the question "How many relevant items are retrieved?".

Precision, also called positive predictive value, is the fraction of relevant instances among the retrieved instances. It answers the question "How many retrieved items are relevant?".

Precision@k is the ratio of correctly identified relevant items within the total recommended items inside the K-long list. Simply put, it shows how many recommended or retrieved items are genuinely relevant.

Recall@k measures the proportion of correctly identified relevant items in the top K recommendations out of the total number of relevant items in the dataset. In simpler terms, it indicates how many relevant items you could successfully find.

Fβ@k combines these two metrics into a single value to provide a balanced assessment. The Beta parameter allows you to adjust the importance given to recall relative to precision.

Together, these metrics quantify the effectiveness of our algorithms and inform the next set of improvement goals.

## 2.2 Knowledge Graph

As has been mentioned in Introduction — Application of automation, that Large Language Models are meant to deliver into big sophisticated systems / large knowledge bases is mainly constrained by LLM context window size. The concept of Knowledge graphs in large knowledge bases is used as a helpful tool in solving LLMs context window size constraints besides the fact that it is also used for structured representation of large knowledge bases for humans. In our specific application, Knowledge Graphs are bridging the gap between raw source code and semantic, architectural and operational understanding [1, 2] without needing human time and attention. In repository mining systems, they enable sophisticated reasoning and retrieval capabilities through graph-based data organization.

Abstract Syntax Tree (AST) parsing provides the foundational layer for repository analysis, decomposing source code into hierarchical representations that capture syntactic relationships and program structure [19, 13]. This language-agnostic approach enables systematic extraction of code entities such as classes, functions, and their interdependencies through static analysis regardless of the technologies used by repository developers.

Vector embeddings encode semantic relationships between code entities as high-dimensional numerical vectors, typically generated through transformer-based models [3, 9]. These dense representations enable similarity computation and clustering of functionally related components within high-dimensional embedding spaces where semantically similar code constructs are positioned in proximity.

We define a Dynamic Knowledge Graph (DKG) as a representation of a software repository's evolving structure and rich topology. It supports fast semantic and relational search over nodes by combining vector embeddings with explicit interdependency edges, enabling efficient retrieval of related artifacts as the codebase changes. Relevant search based on Knowledge Graph could be performed in 3 main approaches:

(1) through embedding-based similarity search where natural language queries are embedded into the same vector space as code entities, enabling retrieval through cosine similarity or other distance metrics in the embedding space [3, 4, 9].

(2) using graph queries where natural language questions are automatically translated into structured query languages (such as Cypher for Neo4j) that traverse the knowledge graph structure directly, leveraging the explicit relationships between nodes [1, 2, 14].

(3) combination of the two above.

This approach creates a Retrieval-Augmented Generation (RAG) architecture where the knowledge graph serves as an external memory system, providing contextually relevant information for code understanding and generation tasks [3, 4, 5].

### 2.2.1 Knowledge Graph Initialization

For the Knowledge Graph initialization, a repository link along with a commit id is needed. Once a repository is checked out, the build graph stage is started. A skeleton of the graph is built by initializing all the Folder and File nodes. All the docstrings and comments are being parsed into the graph along with the raw source code. Technical summary information about the repository is saved into the graph and further serves as core system context for the LLM Agent (from section 4) and an LLM-based summarization generation.

### 2.2.2 Source Code Files Parsing & Metadata Extraction

As next steps
   (1) language-agnostic parsing extracts programming language elements and identifiers Classes, Functions, Member functions etc. and their metadata from source code files with all the relations between them;
   (2) all import/include/reference statements are recorded to form file-to-file reference edges;
   (3) function-level call relationships and class inheritance are extracted where available.
   For each parsed artefact we keep raw content, docstrings and comments, available signatures/types, and metadata input / output types within node properties.

### 2.2.3 LLM-based Summarization & Embedding

Short intent-focused natural-language summaries are generated for Folders, Files, Classes and Functions using compact LLM variants for graph enrichment. Vector embeddings are computed for generated node properties with representative function/class-level contexts. Embeddings are stored within node properties [8, 20].
   LLM Task Objective and Analytical Instructions are matched in LLM prompts from both sides: from user input processing stage and description generation stage (system analytical output).

### 2.2.4 Dynamic Knowledge Graph

Once a graph has been created based on a specific commit id, it could be updated using Update graph Stage. The updated graph stage could be executed based on a new commit id from the same branch as the commit id being used in the load graph stage. The graph that changes with time is Dynamic Knowledge Graph.
   Graph update includes execution of source code parsing, summarization and embedding operations for files changed. For each modified node timestamp along with other node properties is updated. Basically, the update graph stage uses the same functionality as load graph stage, the only one difference — it is also updating the existing nodes in the graph besides the creation of the new ones [1, 16].

### 2.2.5 Graph Objects' Schema

Main object types within our Knowledge graph are Nodes and Edges with their properties.
   Node types:
   (1) **Folder** — path, description, description embedding, basic metadata.
   (2) **File** — path, language, size, last-modified timestamp, description, description embedding, code embedding.
   (3) **Class** — name, docstring, description, description embedding, code embedding.
   (4) **Function** / **MemberFunction** — name, signature, docstring, description, description embedding, code embedding, lines.
   (5) **Root** — name of the repository, description of the repository, scope and technologies.
   Edge types:
   (1) **Contains** connects files, folders, functions and classes.
   (2) **Inherits** connects classes and interfaces.
   (3) **Tests** connects files, functions.
   (4) **Implements** connects files, classes, functions.
   (5) **Calls** connects functions.
   Given objects schema allows our system to reach all the most hidden system dependencies and objects of the repository [13, 10].

### 2.2.6 Knowledge Graph Example of Poetry Repository

For a clearer picture of our Knowledge Graph, we will use enriched Knowledge graph from Poetry source code repository. GitHub link: https://github.com/python-poetry/poetry, commit id: a622badfe8b2f9223c5d1d93f11d89e9cf67d877.
   Objects statistics within the Poetry repository are represented in Table 1.
   The Knowledge Graph of a repository is created for a specific revision and can be updated based on a new revision – which allows us to collect and dynamically operate the information about the changes and development process within the repository. Figure 2 represents an enriched knowledge graph from neo4j browser, generated for Poetry, where folders are represented in blue, files in orange, classes in red, functions in brown, and member functions in green.
   The given graph structure reflects the repository architecture. All the nodes are architecturally and contextually decomposed.

| File types counts: | Nodes: | Edges: |
|---|---|---|
| - Source code files — 280<br>- Documentation / Readme files — 300<br>- Other file types (env, settings files) — 100 | - All nodes: 3245<br>- File: 763<br>- Folder: 280<br>- Function: 1016<br>- Class: 228<br>- MemberFunction: 967 | - All relations: 5055<br>- Contains: 1043<br>- Implements: 2201<br>- Refers: 995<br>- Inherits: 104<br>- Tests: 712 |

Table 1: Statistics of the Poetry repository

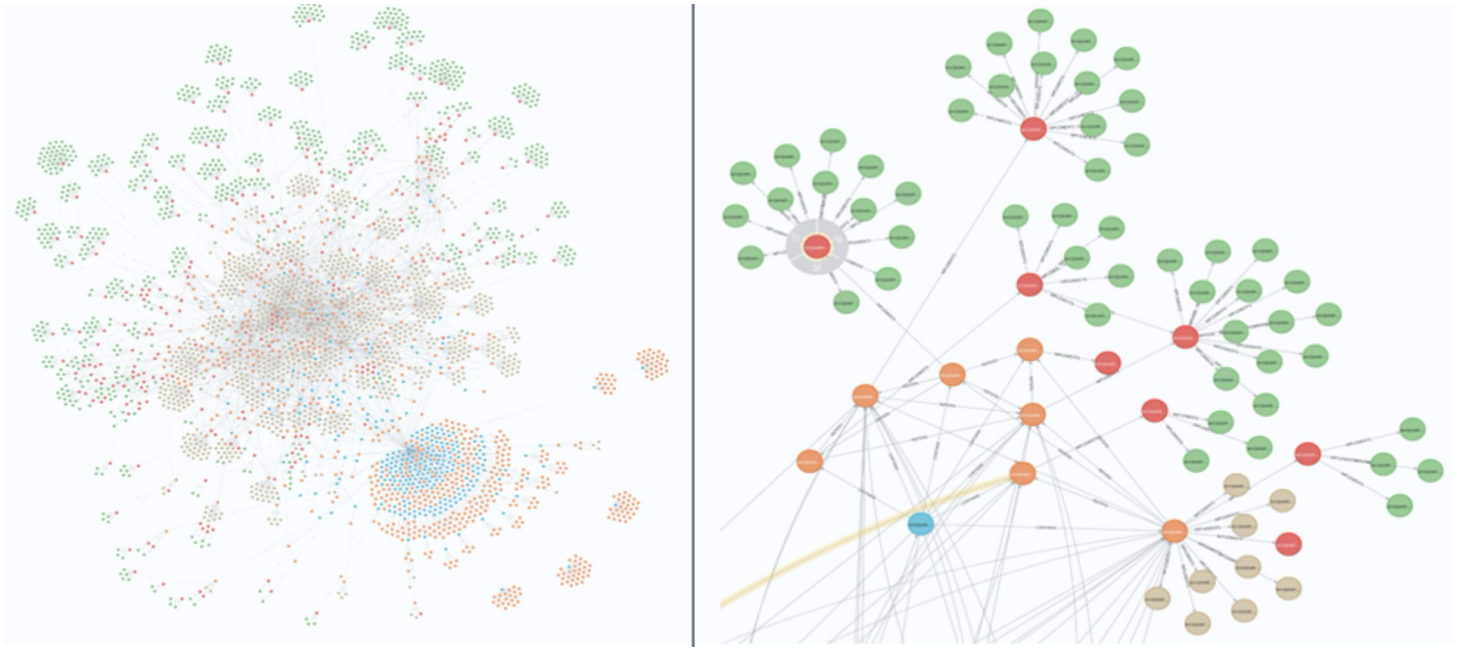

Figure 2: Poetry repository Knowledge Graph

## 2.3 Search and Retrieval

One of the most powerful tools Retrieval Augmented Generation (RAG)-based systems and Knowledge graphs can deliver is Semantic Search among documents. Given search purely relies on Semantic understanding of given documents without any keywords or string matching, pure linear algebra, which greatly powers up knowledge bases & knowledge graph search semantic precision [3, 4, 5].

In our specific RAG application of context mining of large codebases the retrieval objective is to return an ordered set of objects (folders, files, classes, functions, member functions) most likely to be modified to resolve a repository bug / add repository feature / enhance repository or just related to a user query [16, 7].

The obvious application of our RAG system is retrieval of relevant files / folders from repositories looking only at bug titles and descriptions. In order to improve relevant search even more, we use graph relationships with semantic search [2, 14].

### 2.3.1 Search Relevant **Algorithm**

Our Search relevant algorithm is a version of semantic search enriched with some small features, like: Query Preprocessing, Semantic Search, Graph Traversal, LLM discovery. Figure 3 represents a schematic representation of the search relevant algorithm [1, 2].

Input of the algorithm is a NL user query, that can represent bug description, new feature specification or just any question related to a source code repository. All experiments and numbers that we use for stages efficacy evaluation were obtained while experimenting with Apache Airflow repository. The results can differ for different repositories, user input descriptions, and should be tested on a variety of different cases. The results that are represented below can be considered only as examples and preliminary conclusions and are based only on 40 tests cases for the repository.

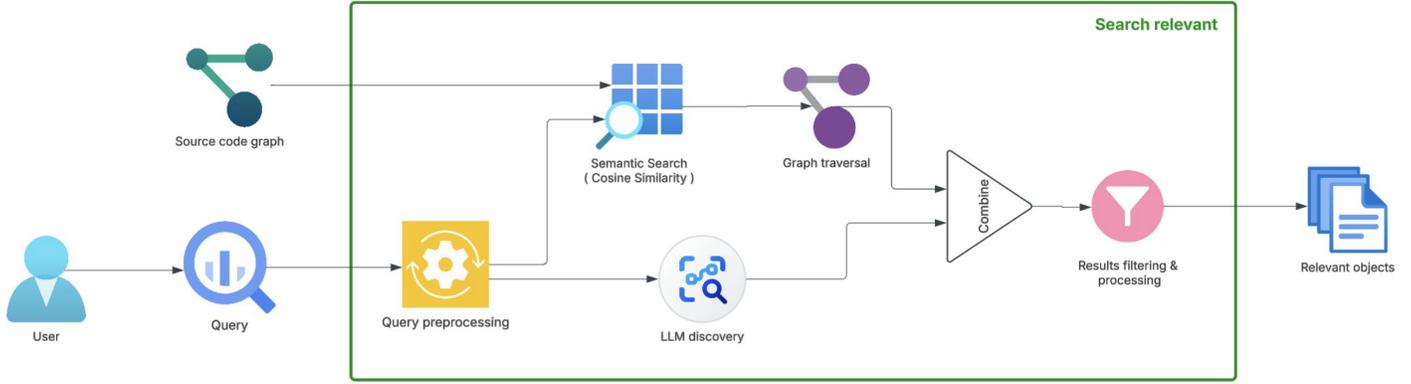

Figure 3: Search relevant algorithm schema

#### 2.3.1.1 Query preprocessing stage

Query preprocessing stage is meant to normalize and preprocess the raw user input, as well as to enrich with relevant information. For a given stage, LLM is used with specific fine-tuned prompt. In our research, we used OpenAI gpt-4.1. We experimented with different approaches of query preprocessing during evaluation:

(1) None — no query preprocessing, we are passing raw input text to the next stage.
(2) LLM — we are preprocessing input text with LLM. Then the answer is passed to the next stage.
(3) Concatenated LLM query — we are preprocessing input text with LLM. Then the answer is concatenated with the raw query and passed to the next stage.
(4) Selective LLM query — we are passing both raw text and answer preprocessed with LLM to the next stage. Then in that stage we are choosing between them based on cosine similarities.

Table 2 shows metrics of different query preprocessing techniques and their efficacy [16, 8].

| Query preprocessing | none | Selective LLM query | LLM | Concatenated LLM query |
|---|---|---|---|---|
| Recall | 0.534 | **0.674** | **0.674** | 0.584 |
| Percentage Found, N % | 5.0 | 5.0 | 5.0 | 5.0 |

Table 2: Query preprocessing techniques' metrics

As you can see, the Query Preprocessing Stage is very important for our algorithm. It gave us a significant gain in our metrics —- up to 0.14 increase in recall, our primary metric.

#### 2.3.1.2 Semantic Search Stage

Semantic search stage is a central stage of the solution that consists of two steps:
(1) Compute embeddings from text obtained from the previous stage (user query/ preprocessed user query).
(2) Perform a semantic search for each node in the graph.

Semantic search means calculating cosine similarity between input embeddings and embeddings of each node in the graph. As a result, we have a ranked list of each node with corresponding similarities. Formula 1 represents cosine similarity of two vectors.

$$\text{similarity}(A, B) = \cos(\theta) = \frac{A \cdot B}{\|A\| \|B\|} = \frac{\sum_{i=1}^{n} A_i B_i}{\sqrt{\sum_{i=1}^{n} A_i^2} \sqrt{\sum_{i=1}^{n} B_i^2}}.$$

**Formula 1:** Cosine similarity of two vectors

One of the goals of our experiments with semantic search stage was to compare different embedding models in terms of metrics. Bert model stands for sentence-transformers/all-MiniLM-L6-v2 model. OpenAI stands for text-embedding-3-small model. BGE_small stands for BAAI/bge-small-en-v1.5 model. BGE_large stands for BAAI/bge-large-en-v1.5 model. Table 3 represents metrics for different embeddings models for our solution [9].

| Embedding model | Bert | BGE_small | OpenAI | BGE_large |
|---|---|---|---|---|
| Recall | 0.613 | 0.638 | **0.697** | 0.678 |
| Percentage Found, N % | 20 | 20 | 20 | 20 |

Table 3: Comparison of different embedding algorithms

As you can see, choosing an embedding model is important. We decided to use BGE_large because it is easier and cheaper to work with and it gave us pretty good results [9].

#### 2.3.1.3 Graph Traversal Stage

Graph traversal stage is meant to enrich the list of files selected by semantic search with closely related and connected files. The given stage allows the system to reach all the most hidden and problematic repository object parts. The current stage is variable and depends on some initial set of control parameters, such as:
(1) Relation types — which type of connections to use to enrich results.
(2) Node types — which labels are allowed during the traversal stage.
(3) Direction — whether it should move from parent to child in a graph or vice versa.
(4) Depth — how deep the results should be enriched by connections.

As a result of our experiments, we decided to traverse from found functions by one "Calls" connection up to another function that was calling it, if it exists. Then, for the found functions, we traverse by the "Implements" relationship up to the files where these functions are written [2, 14].

| Algorithm | Semantic search only | Semantic search with graph traversal |
|---|---|---|
| Recall | 0.528 | 0.534 |
| Percentage Found, N % | 4.8 | 4.8 |

Table 4: Comparison of semantic search with and without graph traversal

As we can see, using DKG traversal slightly improves our metrics.

#### 2.3.1.4 LLM Discovery Stage

The idea of the LLM discovery stage came when we saw that some user input may contain list of files that are related to it. We decided to add a stage that addresses this particular case and allows us to extract from a query all related file names and paths if they are mentioned in the input.

We ask LLM about code, logs and tracebacks etc., mentioned in user input, and ask to generate an answer with files that are relevant to the given issue. We combine files found at this stage with files found using semantic and graph traversal stages. Table 5 represents the result of the stage's performance [16].

| Algorithm | Semantic search with graph traversal | Semantic search with graph traversal and LLM discovery |
|---|---|---|
| Recall | 0.674 | **0.707** |
| Percentage Found, N % | 5.1 | 5 |

Table 5: Comparison with and without LLM discovery

It follows from our results that LLM discovery stage helped us to increase primary metric: recall. All these stages are being selected empirically, meaning that they were added only after evaluation.

### 2.3.2 Search Relevant Metrics

After performing experiments and completing the final version of our algorithm, we evaluated it. The primary metric for evaluation was Median of Recall@k across different test cases of a single repository. As auxiliary metrics we also took Median of Precision@k and Median of Fβ@k across different test cases of a single repository. Languages represent the primary language used in the specified repository. Test cases amount means the number of test cases picked up for the selected repository. Files returned is specified number of files that were returned, files total means the total number of files in a particular repository, percentage of returned files is their fraction.

For evaluation, we decided to take repositories of different sizes and programming languages. Below is the list of repositories we choose for our evaluation session:
(1) React — large web backend (JavaScript, TypeScript)
(2) Poetry — Python modern package manager (Python)
(3) Pytest — testing software (Python)
(4) Eslint — code analysis tool (JavaScript, TypeScript)
(5) Junit-framework — junit testing (Java, Kotlin)

Tables 6 and 7 represent metrics per repository for different K metrics.

| Repository | Languages | Test cases | Files total | Files returned | % Files returned | Median Recall@50 | Median Precision@50 | Median F$\beta$@50 |
|---|---|---|---|---|---|---|---|---|
| Poetry | Python | 48 | 893 | 50 | 5.6% | 1 | 0.02 | 0.927 |
| Pytest | Python | 24 | 610 | 50 | 8.2% | 1 | 0.04 | 0.927 |
| Junit-framework | Java | 42 | 1953 | 50 | 2.6% | 1 | 0.04 | 0.673 |
| React | JavaScript, TypeScript | 34 | 6623 | 50 | 0.8% | 1 | 0.02 | 0.693 |
| Eslint | JavaScript | 60 | 2219 | 50 | 2.3% | 1 | 0.04 | 0.762 |

Table 6: Results for top 50 files

| Repository | Languages | Test cases | Files total | Files returned | % Files returned | Median Recall@10 | Median Precision@10 | Median F$\beta$@10 |
|---|---|---|---|---|---|---|---|---|
| Poetry | Python | 48 | 893 | 10 | 1.1% | 0.5 | 0.1 | 0.453 |
| Pytest | Python | 24 | 610 | 10 | 1.6% | 0.584 | 0.1 | 0.542 |
| Junit-framework | Java | 42 | 1953 | 10 | 0.5% | 0.5 | 0.2 | 0.357 |
| React | JavaScript, TypeScript | 34 | 6623 | 10 | 0.2% | 0.5 | 0.1 | 0.433 |
| Eslint | JavaScript | 60 | 2219 | 10 | 0.5% | 0.5 | 0.1 | 0.433 |

Table 7: Results for top 10 files

Analyzing the results of these two tables, we can see that 50 files usually are enough to find the majority of relevant files and 10 files are usually enough to find 50% of relevant files.

The next question concerns runtime performance: how fast is our solution? We address this by benchmarking across repositories that vary in programming language and scale (size), enabling a comparative assessment under heterogeneous conditions. We measured graph creation time, search relevant algorithm time with and without LLM preprocessing stage. Note that provided numbers were made on m4pro CPU. Graph creation time was measured with using cache of nodes' descriptions and embeddings.

So, as we can see from Table 8, our search relevant algorithm can perform search without LLM up to 2 seconds on big repositories with 16.7k nodes total.

### 2.3.3 Example of Search Relevant on the Poetry Repository

We next present an illustrative case study of our context-aware retrieval over a Knowledge Graph. As the target corpus, we used the Poetry repository https://github.com/python-poetry/poetry; revision: 427d922; issue: #10429; PR: #10431. The issues description is: "Unexpected Behavior: poetry new . Acts Like poetry init Without Creating Project Structure".

The query context is defined by the Issue–PR pair above. Figure 4 visualizes the resulting search-relevance similarity graph; nodes correspond to files, and edges encode similarity within the KG. Files ranked as most relevant by the algorithm are highlighted in green.

On the given graph representation all the nodes from Poetry repository are displayed, nodes, different type have different sizes, order of node types by sizes: Folder, File, Class, Member Function, Function. All the nodes are in colour range from orange to green, based on cosine similarity with preprocessed user issue information, the greener - the higher similarity, names of top 20 nodes are displayed with their labels.

| Repository | Languages | Nodes total | Relationships total | Graph creation time with cache, s | Query time with LLM, s | Query time without LLM, s |
|---|---|---|---|---|---|---|
| Itsdangerous | Python | 188 | 592 | 2 | 13 | 0.30 |
| Flask | Python | 1,578 | 5,983 | 25 | 10 | 0.40 |
| Poetry | Python | 4,085 | 17,081 | 206 | 9 | 0.78 |
| Pytest | Python | 6,663 | 28,144 | 198 | 13 | 0.59 |
| Okhttp | Kotlin | 1,288 | 2,537 | 22 | 10 | 0.74 |
| Junit-framework | Java | 16,731 | 165,231 | 1,208 | 10 | 1.93 |
| React | JavaScript, TypeScript | 18,499 | 425,141 | 2,832 | 19 | 5 |
| Eslint | JavaScript | 4,460 | 8,788 | 388 | 18 | 2 |

Table 8: Performance and latency

For evaluation, we take the PR's modified files as ground truth. If we look in the PR itself, we will see 4 files and from those 4 files in total 6 objects have been modified, all of those objects are in top-20 results returned by our Search Relevant algorithm. Modified files and objects from a resulting PR on GitHub:

(1) src/poetry/console/commands/new.py;
(2) tests/console/commands/test_init.py;
(3) tests/console/commands/test_new.py;
(4) src/poetry/console/commands/init.py/ InitCommand;
(5) src/poetry/console/commands/new.py/ NewCommand;
(6) src/poetry/console/commands/new.py.

Within a candidate set of k = 20 files, the method retrieved all relevant files, yielding Recall@20 = 1.00 and Precision@20 = 0.20. These results indicate that the approach can recover the complete set of relevant artifacts while confining review to a compact candidate pool; we view this as an encouraging empirical demonstration of the algorithm's effectiveness.

## 2.4 Repository Clustering

We also use clustering to help find similar files inside a repository. The idea is simple: files that change together are usually related—by meaning or by dependencies. We group files into clusters using embeddings and known links (e.g., imports, co-change). During retrieval, we prioritize files from the same cluster as the query context, which increases the chance of returning relevant results.

To make repositories easier to understand and search, the system builds a multi-view representation for each file that combines structural signals (connectivity, centrality), semantic signals (LLM-based embeddings), and temporal signals (commit history). We then apply three complementary clustering methods—semantic (embedding-based), Louvain, and label propagation—where the first captures similarity in meaning and the latter two exploit network topology to reveal community structure. Integrating these views and methods improves user understanding of the codebase and supports better personalization and domain-specific retrieval [1, 13].

Clusters may be identified by numeric IDs as well as by human-readable labels generated by LLMs. Numeric IDs are typically sufficient for enriching search results (e.g., by adding files from the same cluster), while textual labels are especially useful for visualization.

The visualization service can display cluster meta-nodes, supports drill-down into cluster details (file nodes, call edges, summaries), and allows issue-specific subgraph views that highlight candidate files along with their immediate neighbors and cluster memberships. Interactive filters enable users to narrow views by language, recency, or label. Clicking on a node surfaces its summary, representative snippets, and connected nodes. For examples of clustering, we will use Flask repository https://github.com/pallets/flask with 83 source code files in it.

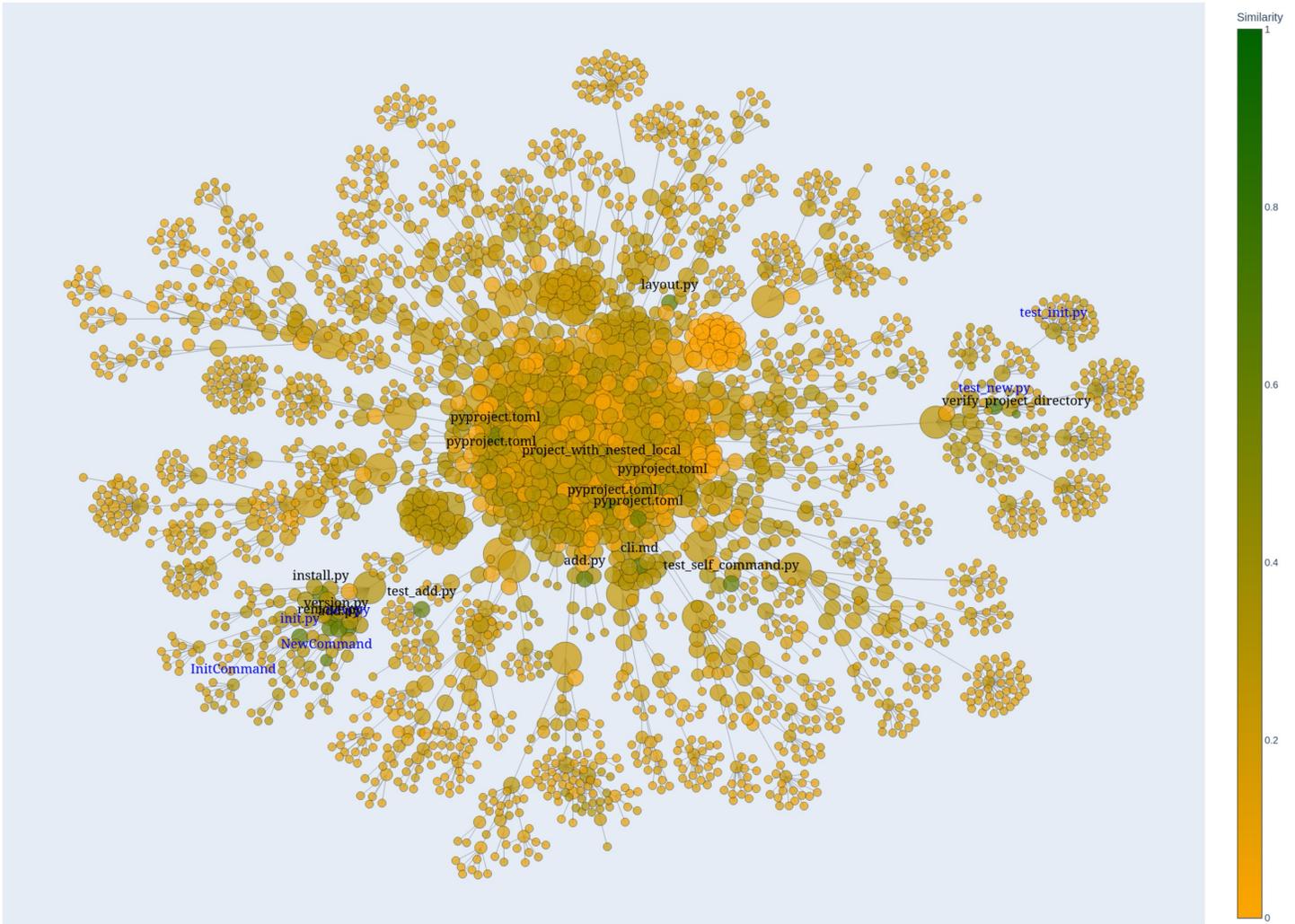

Figure 4: Search Relevant over Poetry repository

### 2.4.1 Semantic Clustering

Semantic clustering groups elements based on similarities in their semantic representations, typically derived from embeddings of textual or behavioral features. This aligns clusters with latent meaning and user queries. As is common in text embedding workflows, dimensionality reduction with Uniform Manifold Approximation and Projection (UMAP) is applied.

The system generates multiple different cluster groups, which are then ranged, sorted and compared, until the most "relevant" of them is selected. A whole algorithm is developed for selecting the best clustering. We delete too fractional or too general clustering results, ones with too many unsorted files, and so on. Here and below, an example of clustering for Flask repository is shown in Figure 5.

The repository consists of 83 files, that have been split into 16 clusters with corresponding labels and cluster size represented in the figures below.

### 2.4.2 Clustering Comparison

Louvain clustering, a graph-based community detection method, optimizes modularity by iteratively merging nodes into densely connected communities, providing scalability and robust performance for large networks. At each step, individual nodes making up a community are combined into just one node, while the edges between the new nodes are (re)assigned a weight based on the weights of the edges inside the old communities. Unclustered files — and, optionally, clusters with just two items — are afterwards grouped together to form the 'misc' cluster.

The Flask repository has been split into 16 clusters with corresponding labels and cluster size shown in the figures above.

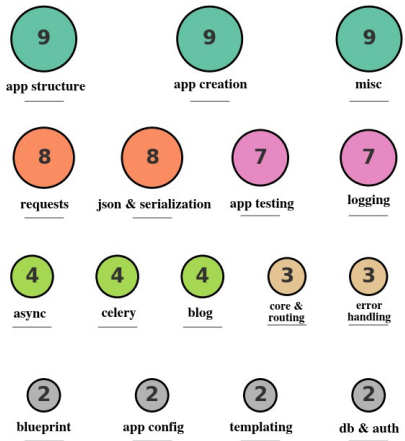
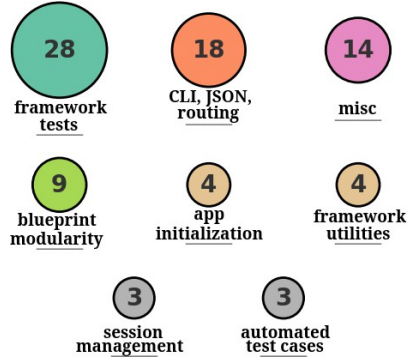
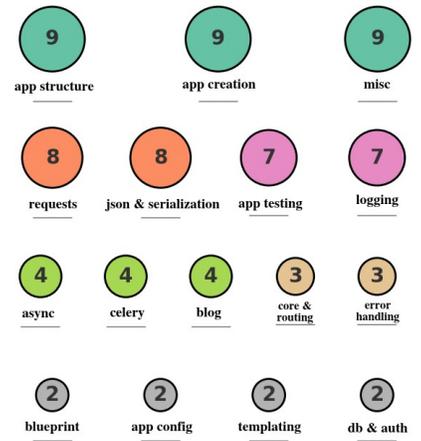

Figure 5: Semantic clustering example

Figure 6: Louvain clustering example

Figure 7: Label propagation clustering example

### 2.4.3 Label Propagation Clustering

Label propagation uses an iterative diffusion process in which nodes adopt the most frequent label among their neighbors, enabling efficient, parameter-free community discovery. Still, this clustering algorithm seems to be the least informative of all three, often reducing half of the nodes to just one cluster and/or leaving many unclassified nodes, with little possibility of adjusting its behavior. Another problem of this method is extensive use of random choices, which leads to different results on the same data.

The Flask repository has been split into 16 clusters with corresponding labels and cluster size represented in Figure 8.

## 3 Application of Dynamic Knowledge Graph

As part of our research, we integrated the DKG into an agent-based coding assistant and conducted an empirical evaluation of its usefulness in day-to-day engineering workflows. In controlled trials on real repositories, we observed consistent improvements on common maintenance tasks—most notably impact analysis (identifying the blast radius of a change), bug localization (narrowing suspect files/components), and architecture conformance (detecting drift from intended module boundaries)—as well as related activities such as test selection and dependency inspection. While we do not report quantitative benchmarks here, practitioner feedback and qualitative time-to-insight measures indicated that the assistant shortened investigative loops and reduced handoffs.

Conceptually, LLM-based agents operate over the DKG of the source-code repository, which encodes files, tests, build artifacts, developers, and their temporal/structural relations. This pairing yields a safe and interpretable natural-language interface to repository knowledge: developers can pose questions in plain language, and the agent compiles those intents into deterministic, read-only graph queries against a typed schema. This workflow is convenient for practitioners because it reduces cognitive load and context switching across the IDE, VCS, CI, and issue tracker; answers are returned with provenance (node summaries, code snippets, and links to commits/issues), enabling quick auditability and handoff to code review [14, 6].

Agents are particularly effective over a DKG because the graph exposes a declarative, provenance-rich query surface (e.g., Cypher) whose deterministic semantics constrain hallucination and ground responses in concrete artifacts. The schema aligns with developer concepts—modules, dependencies, ownership, coverage, commit history—while temporal edges capture co-change and regression patterns that are essential for maintenance reasoning. Under conservative constraints, the agent maps user intent to read-only subgraph retrieval and evidence-first summarization: results are pruned for readability, accompanied by concise, human-oriented explanations, and cross-referenced with verifiable sources (provenance links). This design preserves safety (no repository mutations), enhances explainability, and supports independent validation by engineers [1, 16, 15].

## 4 Limitations and Ethical Considerations

Staleness is a fundamental concern: graphs represent snapshots of code at particular commits, so alignment to commit hashes is required when applying the system in CI or historical analysis. Importantly, the pipeline supports incremental updates so that new commits, modifications, and deletions can be parsed and the graph updated to reduce structural drift. But still, the main challenge in using a DKG in real-world settings is freshness at scale. The graph must be kept continuously up to date, and at realistic sizes it can be computationally heavy—whether built on a local machine or hosted on a server. In a server-centric architecture, maintaining versioned views for multiple repository revisions is particularly tricky and costly, because different developers may work on different commits/branches concurrently.

Heterogeneity across programming languages adds further complexity. Features such as dynamic dispatch, reflection, macros, code generation, and runtime imports make many dependencies difficult to recover via static analysis alone. Dynamic analysis—instrumentation, tracing, and coverage collected at runtime—is harder to deploy and scale, but it reveals the dependencies exercised during execution, their strength/frequency, and latent or indirect couplings and temporal trends that static methods often miss [7, 13, 19].

Another key limitation is human factor: user-provided issue descriptions are often incomplete, inconsistent, or ambiguous. Issues may omit critical context, combine multiple problems in a single thread, provide trimmed stack traces or partial logs, or use vague terminology. These human-data characteristics reduce signal quality for both semantic matching and intent extraction. The system applies normalization and conservative intent-extraction rules to mitigate noise, but inconsistent or insufficient issue descriptions remain an important source of error and must be considered in evaluation design and interpretation of retrieval performance [16].

From a safety standpoint, the tool focuses on retrieval and explanation and avoids proposing automatic code modifications; human review is required before any change is applied.

To reduce risk and improve trust, we adopt several conservative practices. Generated summaries are labeled clearly, for example "LLM-suggested", and raw evidence such as code snippets and provenance is surfaced alongside recommendations. Expansion depth and result sizes use conservative defaults to limit spurious context. Agent actions and queries are logged for auditability and post-hoc analysis.

# 5 Discussion and Future Work

Future work spans several complementary directions aimed at improving both retrieval quality, architectural insight and addressing some limitation of the algorithm.

(1) We will incorporate dynamic signals that reflect the program as executed: dynamic call graphs/trees (per test and aggregated), runtime object/receiver types and dispatch targets (capturing real type–dependency edges), and actual control-flow paths taken (including branch/feature-flag activation, error paths, and path profiles). These observations induce runtime edges and weights (e.g., call frequency, path prevalence, latency/exception rates) and annotate nodes with empirical usage statistics. By fusing these dynamic relations with the static graph, the system can localize fault-relevant files and components using evidence from concrete executions, rather than relying solely on static structure [7].

(2) We will construct temporal graphs over the commit history to model co-change patterns and release rhythms. A time-aware graph (with dated edges and commit-level provenance) should improve cluster coherence by distinguishing stable architectural couplings from transient, release-specific edits; it also enables link-prediction and drift detection across versions.

(3) We will train supervised re-rankers on historical issue→PR pairs to capture how engineers have resolved similar problems in the past. These learning-to-rank models can exploit heterogeneous features—textual similarity between issue descriptions and diffs, developer ownership and expertise, co-change frequency, test impact, and dependency distance—to prioritize candidate files and subgraphs for investigation [16].

(4) We will enrich the graph with heterogeneous signals from pull-request metadata (reviews, labels, merge outcomes), commit messages/diffs, issue-tracking artifacts (status transitions, components, severity), deployment/CI scripts (pipelines, artifacts, environments), and documented architectural patterns. This broader evidence base should help uncover latent dependencies (implicit couplings) and emergent clusters, supporting tasks such as architecture conformance checks, change-impact analysis, and proactive regression monitoring [6, 1].

Figure 8 illustrates a Knowledge Graph that integrates multiple facets of the project—source files and modules; tests and coverage; build artifacts and pipelines; developers and code ownership; services, configs, and environments—linked by typed, provenance-rich edges (imports, calls, co-change, verifies, deploys-to, owns). As this graph is incrementally enriched, it can serve as a governed "source of truth" for the repository: each node/edge carries versioning metadata and timestamps; provenance ties back to commits, PRs, and CI runs; and freshness policies ensure that derived views (e.g., clusters, ownership maps) remain consistent with the evolving codebase. In this role, the DKG becomes a stable substrate for both agentic querying and analytical workloads (impact reports, dependency risk scoring, test-selection strategies) [1].

Finally, we hypothesize that developer-in-the-loop workflows can further improve outcomes. By capturing lightweight human feedback signals—accept/reject on suggested files, labels for cluster membership, confirmations of architectural intent, and edits to generated rationales—we can drive iterative refinement of cluster boundaries and retrieval behavior (e.g., via active learning or weak-supervision schemes). We will evaluate these loops with both offline metrics (e.g., precision@k for localization, NMI/ARI for clustering) and online measures (time-to-insight, fix success rate), with the goal of steadily improving the assistant's recommendations while preserving safety, provenance, and interpretability.

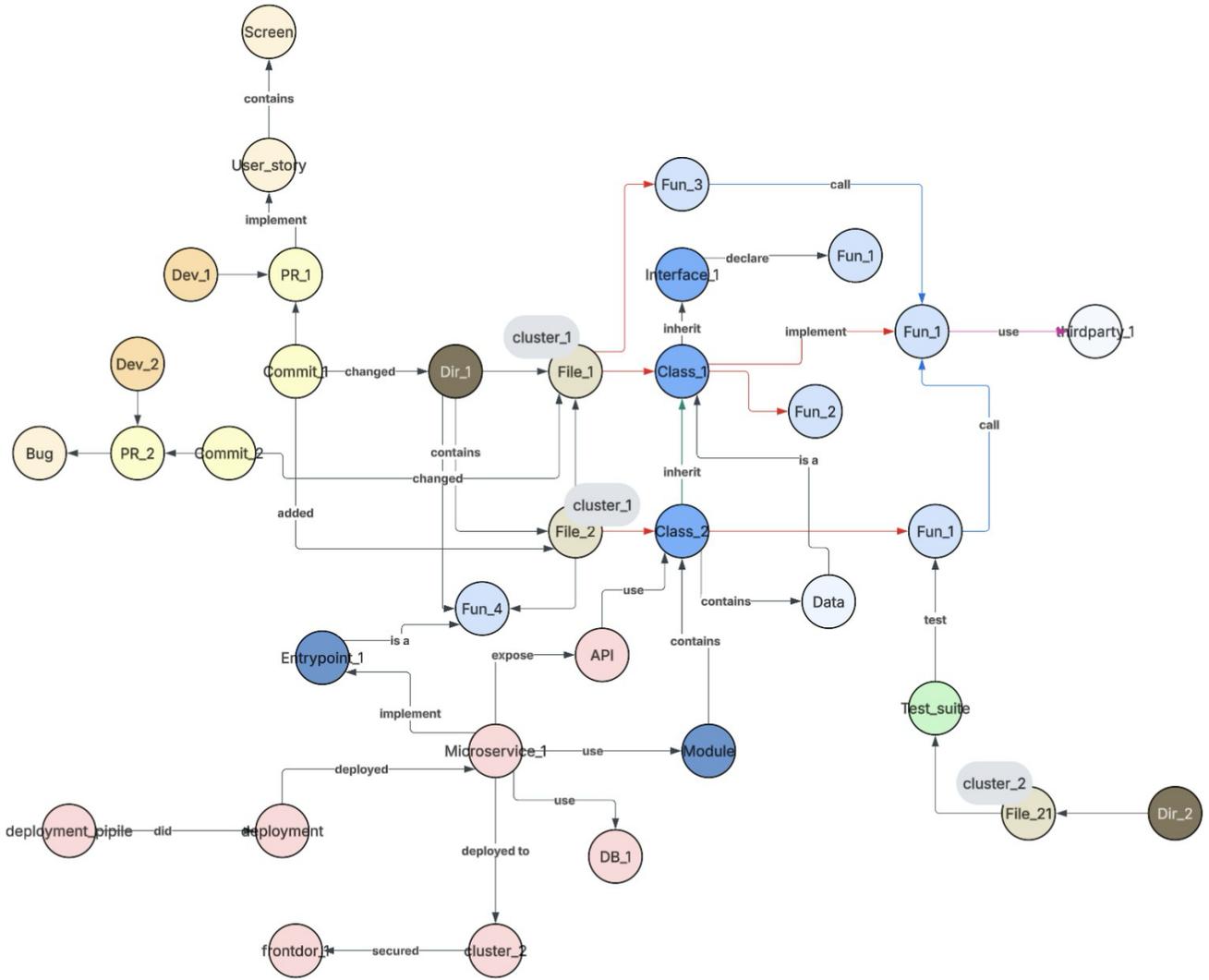

Figure 8: Rich Knowledge Graph

## 6   Conclusion

We present a practical pipeline that converts software repositories into vectorized knowledge graphs and uses these graphs—together with semantic embeddings and LLM-based summarization—to retrieve files relevant to natural-language queries. The system emphasizes explainability and reproducible evaluation, using GitHub issues/PRs as ground truth. We evaluate with median Recall@k (primary), Precision@k, and F$\beta$@k across test cases per repository. In A/B experiments, some techniques produced measurable gains, while others had negligible effect. Future work includes enriching the graph with additional semantic and temporal signals (e.g., PR metadata, commit history, runtime traces), refining algorithms and prompts, and targeting robust performance such as Recall@30 ≥ 0.90 across repositories, independent of repository size [3].